\documentclass[aps,floatfix,twocolumn,showpacs,pra]{revtex4}
\usepackage{graphicx,bm}
\usepackage{amsmath}
\usepackage{amssymb}
\usepackage{amsfonts}%
\usepackage{graphicx}
\usepackage{color}

\usepackage[T1]{fontenc}
\usepackage[latin2]{inputenc}
\usepackage[english]{babel}

\usepackage{bm}

\newcommand{\meanv}[1]{\left\langle #1 \right\rangle}
\newcommand{\bb}[1]{\left( #1 \right)}

\newcommand{\be}{\begin{equation}}
\newcommand{\ee}{\end{equation}}
\newcommand{\bea}{\begin{eqnarray}}
\newcommand{\eea}{\end{eqnarray}}

\newcommand{\wT}{\omega_{\perp}}

\begin{document}
\title{Dipolar dark solitons}
\author{Krzysztof Paw\l owski}
\affiliation{%
Center for Theoretical Physics, Polish Academy of Sciences, \\
Al. Lotnik\'ow 32/46, 02-668 Warsaw, Poland
}
\author{Kazimierz Rz\k a\.zewski}
\affiliation{%
Center for Theoretical Physics, Polish Academy of Sciences, \\
Al. Lotnik\'ow 32/46, 02-668 Warsaw, Poland
}

\begin{abstract}
We numerically generate, and then study the basic properties of dark soliton-like excitations in a dipolar gas confined in a quasi one dimensional trap. 
These excitations, although very similar to dark solitons in a gas with contact interaction, interact with each other and can form bound states.
During collisions these dipolar solitons emit phonons, loosing energy, but accelerating. 
Even after thousands of subsequent collisions they survive as  gray solitons and finally  reach dynamical equilibrium with background quasiparticles.
Finally, in the frame of classical field approximation, we verified, that these solitons appear spontaneously in thermal samples, 
analogously to the type II excitations in a gas of atoms with contact interaction.
\end{abstract}


\maketitle
\section{Introduction}
The main motivation to our work comes from recent progress concerning the role of dark solitons in the physics of the gas with contact interaction only.
First, let us remind, that solitons are localized waves propagating through non-linear media without any spreading 
and also moving through each other without changing the shape.
Among equations of broad interest for physicists, there are a few which exhibit solitons. 
One of them is the non-linear one dimensional Schr\"odinger equation:
        \begin{equation}
	        i\hbar \partial_t \, \psi (z) = \bb{-\frac{\hbar^2 \partial^2_z}{2 m} + g |\psi(z) |^2 } \psi (z),
	        \label{eq:1dGPE-contact}
        \end{equation}
in which one can find bright (the case with attracting forces, $g<0$) and dark (the repulsive case $g>0$) solitons.

The Eq. \eqref{eq:1dGPE-contact} applies, among others, to the dynamics of bosons interacting via a $\delta$-like potential, i.e. the  interaction potential between two bosons $V(z-z^{\prime})$ is equal to $g\delta (z-z^{\prime})$.
In this context, the Eq. \eqref{eq:1dGPE-contact} is only an approximation valid in the variously defined weak interaction limit
\cite{Lieb2005}. 
Such a gas of bosons, can be described by more fundamental many-body model, the Lieb-Liniger model, which has been used to find rigorously the ground state \cite{Lieb1963} as well as the elementary excitations \cite{lieb1963II}.
It turns out, that there are two branches of elementary excitations, staying on the same footing.
The first branch  has been identified as the Bogoliubov excitations.
The second one, consisting of the so called "type II" excitations, within research performed by the next generations of physicists turned out to be solitonic branch \cite{kulish1976,kolomensky2000,jackson2002,carr2010,karpiuk2015,sacha2015}.
It is now clear that the dark solitons, as the elementary excitations, appear spontaneously in a gas at finite temperature \cite{karpiuk2012}.

The equation \eqref{eq:1dGPE-contact}, is not capturing many important and common situations, even in the weakly interacting limit. For example, it is neglecting possible external traps in which the gas is confined, and it is assuming contact interaction only.
In particular, it misses the gas with a dipolar interaction, which is recently studied in more and more laboratories \cite{chromium, erbium,
dysprosium, Ni2008}.
Among many motivations to study the dipolar gas is definitely its unusual excitation spectrum, which in certain situations, close to instabilities,  can have local minimum for non-zero quasi momenta, so called roton spectrum \cite{santos2003}.
It seems natural to ask if here is a place for the second branch and from what excitations it can be build.
Up to our knowledge, there is no study of the elementary excitation spectrum beyond the Bogoliubov approximation, neither 
by solving the underlying many body problem, nor on the mean field level.

Moreover, although the literature about the bright solitons is very reach (see parts of the review \cite{Lahaye2009}), relatively little is known about dark solitons.
For the time being the shapes of dipolar solitons in a gas with both contact and dipolar interaction has been presented \cite{adhikari2014, cuevas2009}. Surprisingly, the life time of the dark solitons can be increased, with some tricks even in the higher dimensional space \cite{nath2008}.
The dark solitons in non-local nonlinear media were the subject of research in optics. 
It has been found that two optical dark solitons interact with each other via non-trivial potential, they can even form a bound state \cite{krolikowski2010}. In this context, although the effective forces between photons can be non-local, it is always assumed that they are not long-range.

Within this work, limited to the mean field description, we look closer at the basic properties of the dark solitons in the dipolar media.
The aim of this paper is to study dark dipolar solitons and contrast them with the true solitons existing in the gas with contact interaction only. 
The analysis is performed within the mean field description, but we include also the finite temperature case.
We start with the detailed description of our model in Sec. \ref{sec:model}. 
After presentation of the basic features of single soliton (shapes, dynamics) in Sec \ref{sec:1soliton}, we discuss a motion of two dipolar solitons in Sec. \ref{sec:2solitons}. Sec. \ref{sec:thermal} is devoted to the analysis of the thermal samples.
\section{The Model \label{sec:model}}
\begin{figure}[htb]
\centerline{\includegraphics[width=8.7cm,clip]{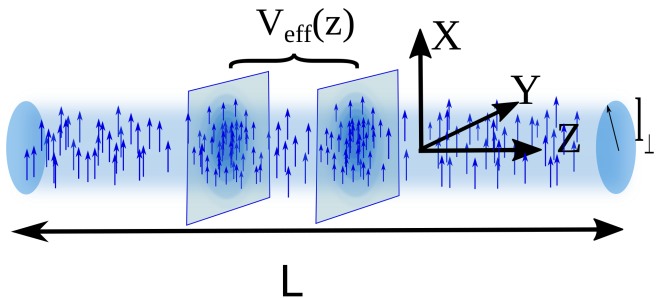}}
\caption{(Color online)
Configuration: Trapping potential is given by a steep harmonic trap in the $X$ and the $Y$ directions (called transverse directions), and a long box with periodic boundary conditions in the $Z$ direction.
The dipole moments of the constituent atoms are polarized along $X$ axis, such that they mostly repel each other.
 \label{fig:Fig1}}
\end{figure}
We assume that the dipolar gas is confined in the potential 
	\begin{equation}
		U(x, y, z) = \frac{1}{2} m \wT^2 \bb{x^2+y^2},
	\end{equation}
with a tight harmonic trap in the $X$ and the $Y$ directions. 
The space in the $Z$ direction is assumed to be finite, with the length $L$ and  with periodic boundary conditions.
These conditions impose a quantization of the momenta $k_z$ in this direction, $k_z$ is a multiple of $\frac{2\pi}{L} $. 

Throughout the paper we use the box units, where $L$, $mL^2/\hbar$, $\hbar^2/mL^2$  and $\hbar^2/mL^2k_{\rm B}$ are the units of length, time, energy and temperature, respectively.
We assume that the dipole moments of the constituent atoms are polarized along the $X$ axis , namely when the dipolar potential in the momentum space has the form
\footnote{We give the dipolar interaction potentials in the momentum space because this form is more useful in numerical treatment.}
	\begin{equation}
		\tilde{V}_{\rm dip} = \frac{g_{\rm dd}^{\rm 3D}}{\bb{2\pi}^{3/2}} \bb{\frac{3k_x^2}{k^2}-1},
	\end{equation}
where $g_{\rm dd}^{\rm 3D}$ is a dipolar coupling strength.
Thus, the atoms move on the circumference of a circle, having the dipole moments perpendicular to the circle-plane.
In this configuration dipoles mostly repel each other, what favors the dark solitons.
The geometry of our problem is sketched in Fig. \ref{fig:Fig1}.

We will investigate the system within the mean field model. The time-dependent mean field wave-function obeys the three dimensional Gross-Pitaevskii equation (GPE)
	\begin{eqnarray*}
		i\hbar \, \partial_t \psi_{\rm 3D} &=& (-\frac{\hbar^2 \Delta}{2 m} + U(x, y, z) + g N |\psi_{\rm 3D} |^2 \\
		&+& \int dr^{\prime} |\psi_{\rm 3D}(\bm{r}^{\prime}) |^2  {\rm V}_{\rm dip} (\bm{r}-\bm{r}^{\prime}) ) \psi_{\rm 3D} (\bm{r}).
	\end{eqnarray*}
We assume that the transverse confinement is so tight, that the wave-function stays in the lowest energy level in the $X$ and $Y$ directions, namely when the following single-mode approximation holds:
	\begin{equation}
		\psi_{\rm 3D}\bb{x, y, z; t} = e^{-i \wT t} \phi_0 (x)\phi_0 (y)\psi (z, t),
	\end{equation}
where $\phi_0$ is the Gaussian wave function:
	\begin{equation}
		\phi_0 (x) = \bb{\frac{1}{\pi l_{\perp}^2}}^{1/4} \exp\bb{-x^2/ (2l_{\perp}^2)},
	\end{equation}
where $l_{\perp} = \sqrt{\hbar/(m \wT)}$ is the transverse oscillator width.
The necessary condition underlying this approximation is that both, the thermal energy $k_{\rm B} T$ and the interaction energy, are much below
the energy of the first excited state in the transverse direction. 
	\begin{equation}
		k_B T, gn \ll \hbar \wT.
	\end{equation}
Assuming such a situation we multiply the three dimensional GPE by $\phi_0 (x)\phi_0 (y)$, integrate the result with respect to
$x$ and $y$ and finally we reach the one dimensional effective GPE
        \begin{equation}
        i\hbar \partial_t \psi = \bb{-\frac{\hbar^2 \partial^2_z}{2 m}
        + N\int dz^{\prime} \,V_{\rm eff} (z^{\prime})\,|\psi(z) |^2 } \psi (z)
	\label{eq:1dGPE}
        \end{equation}
The Fourier transform of the effective one dimensional dipolar potential $\tilde{V}_{\rm eff}(k)$ reads \cite{nath2008, cai2010}.
        \begin{equation}
        \tilde{V}_{\rm eff} (k) = g_{\rm con}  +  3 g_{\rm dd} \bb{ 1 +  f\bb{( \lambda k L)^2 /2} },
	\label{eq:veff}
        \end{equation}
where the constant term with $g_{\rm con} \equiv  \frac{(g - g_{dd}^{3D})}{4\pi l_{\perp}^2}$ mimics a contact interaction, we introduced 
one dimensional dipolar coupling strength $g_{\rm dd} = \frac{g_{dd}^{3D}}{4 \pi l_{\perp}^2}$  and $\lambda \equiv l_{\perp}/L$ is
equal to the aspect ratio of the trap.
The function $f$ which appears in Eq. \eqref{eq:veff} is equal to $f(a) = a e^a \text{Ei}(a)$, where $\text{Ei}$ is the exponential integral.
In this paper we aim at studying the role of pure dipolar interactions only, hence from now on, we will assume $g_{\rm con} = 0$ \footnote{This can be done via Feshbach resonances, setting  $g = g_{dd}^{3D}$}.

\begin{figure}[htb]
	\includegraphics[width=8.7cm,clip]{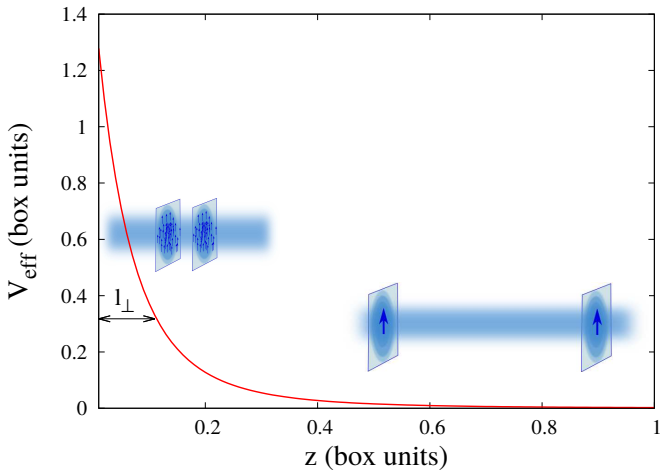}
	\caption{(Color online)
	Effective potential $V_{\rm eff}$ defined in \eqref{eq:veff} in position representation.
	For large distances $z$ it behaves like $1/z^3$, its characteristic range (width at half maximum) is of the order of $l_{\perp}$.
	Parameters: $g_{\rm con} = 0$ and  $g_{\rm dd}=1$.
	 \label{fig:Fig2}}
	\end{figure}

The spatial representation 
 of this potential is visualized in Fig. \ref{fig:Fig2}.
It is worth stressing that Eq. \eqref{eq:1dGPE} describes not a movement of single dipoles, but rather the movement of slices of the dipolar gas as marked in Figs. \ref{fig:Fig1} and \ref{fig:Fig2}. 
If such slices are far from each other, then the details of the distribution of dipoles within such slices plays no role, and we can treat them as single large dipoles. In such situations the interaction between them scales as  $1/z^3$. 
The situation changes, when the distance between them is of the order of the transverse width $l_{\perp}$. This gives the characteristic range of the effective 1D potential \eqref{eq:veff}.
Finally in the limit of slices going to the same place, the effective dipolar potential converges to a constant.

\section{Single solitons \label{sec:1soliton}}
\begin{figure}[htb]
	\includegraphics[width=8.7cm,clip]{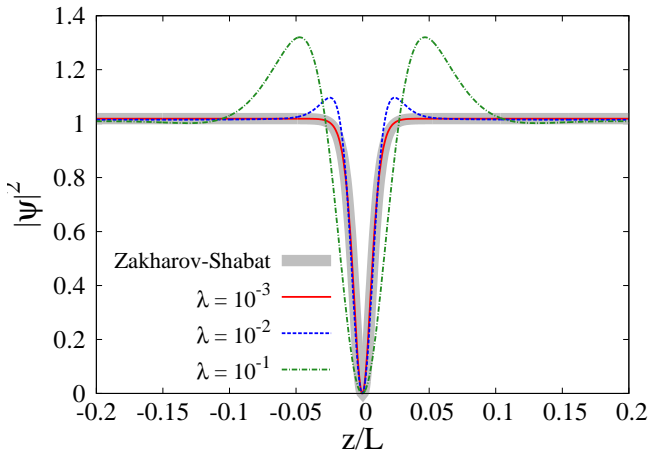}

	\caption{(Color online)
	Shape of the solitonic-like excitations in the dipolar gas. 
	Results of the imaginary time evolution with the phase constraint given in Eq. \eqref{eq:phase1} are marked with: red solid line for  $\lambda = 10^{-3}$, dashed blue for $\lambda=10^{-2}$ and dot-dashed green for $\lambda=10^{-1}$.
	In all figures: $N g_{\rm dd} = 4000$ and $g_{\rm con} = 0$.
	 Gray thick line indicates the Shabat-Zakharov dark soliton \cite{Zakharov1973} for the gas with contact interaction only, $N g_{\rm con} = 12000$.
	The blue dashed line corresponds to experimental-like parameters: the mass and the dipole moment of Dysprosium, assuming $N=600$ atoms and the transverse trap equal to $\wT = 2\pi \times 5$kHz.
	 \label{fig:Fig3}}
	\end{figure}
In the gas with contact interaction only, the phase of a solitonic wave function is changing by $\pi$ around the region with minimal density.
This motivates us to look for the solitons in the dipolar gas within the wave function which minimizes the energy but under constraint of a jump of the phase equal to $\pi$. 
If we only force the jump of the phase equal to $\pi$, then the periodic boundary condition would not be satisfied.
Hence, except the $\pi$ jump required by the soliton, we add to the constraint on the phase a linearly changing function of the position:
	\begin{equation}
		\varphi_1(z) \equiv \frac{\pi}{2} \bb{\frac{2z}{L} - {\rm sign} (z)},
		\label{eq:phase1}
	\end{equation}
such that the phases at the edges of the box are equal. Due to this additional term the gas is moving with the speed $\pi/L$.
When showing the result for the dynamics, we would unrotate this motion.

To find the state with the lowest energy but with the phase equal to the constraint \eqref{eq:phase1} we use the method of the imaginary time evolution, forcing at each step the phase \eqref{eq:phase1}.
Densities computed with this method for three different aspect ratios $\lambda=10^{-3}$, $10^{-2}$ and $10^{-1}$ are shown in Fig. \ref{fig:Fig3}. 
We compare these profiles with the density of a dark soliton from purely contact interacting case \cite{Zakharov1973} marked in Fig \ref{fig:Fig3} with the thick gray line.
As in both cases, a contact interacting gas and the considered here quasi$1$D dipolar gas, the lowest lying excitations are phonons, then comparing their speed of sound   we translate the dipolar coupling coefficient $g_{\rm dd}$ into an effective coupling coefficient $g_{\rm eff}$.
We find $g_{\rm eff} = 3 g_{dd}$. The speed of sound in the dipolar media is equal to $\sqrt{3 n g_{\rm dd}/m}$. The width of the  Shabat-Zakharov soliton is proportional to the healing length, $\xi = \frac{\hbar}{\sqrt{n N g_{\rm eff}/L}} =  \frac{\hbar}{\sqrt{3 n N g_{\rm dd}/L}} = \frac{1}{\sqrt{n a_{\rm eff}}} l_{\perp}$, where $a_{\rm eff}$
is the effective scattering length defined by the relation $g_{\rm eff} = \frac{4\pi \hbar^2 a_{\rm eff}}{m}$.
The Gross-Pitaevskii equation is supposed to be valid in the weakly interacting limit, namely when $ n a_{\rm eff} \ll 1 $, what implies
$\xi \ll l_{\perp}$. 

One can see that the more elongated is the trap, the closer is the result to the soliton in contact interacting gas, marked in Fig. \ref{fig:Fig3} with the thick gray line.
To understand this observation mathematically we look at the system in the momentum space.
The range of the effective dipolar potential in the momentum space is equal to $1/l_{\perp}$, equal in the box units to $1/\lambda$.
Hence, in the limit of a very elongated trap, $\lambda \to 0$ the range of the effective potential is going to $\infty$, and the evolution is dominated by its values relatively close to $k=0$.
The derivative of the effective potential Eq. \eqref{eq:veff} at $k=0$ vanishes, what means that the potential is flat at the origin.
Thus locally, where this 'local' region spans even to high momenta, the interaction potential looks like a constant function, as it would be in the contact interacting case.
The ratio  between the typical width of the soliton and the range of the potential , written in the momentum space and in the box units is equal to $\frac{l_{\perp}}{\xi} = \sqrt{n g_{\rm dd}} \lambda$. 
Thus if the trap is very elongated  and the coupling strength is small,  we expect that the dipolar soliton will be close to the  Shabat-Zakharov solution and moreover that our system will evolve similarly to the gas with contact interactions only.
	\begin{figure}[htb]
	\includegraphics[width=8.7cm,clip]{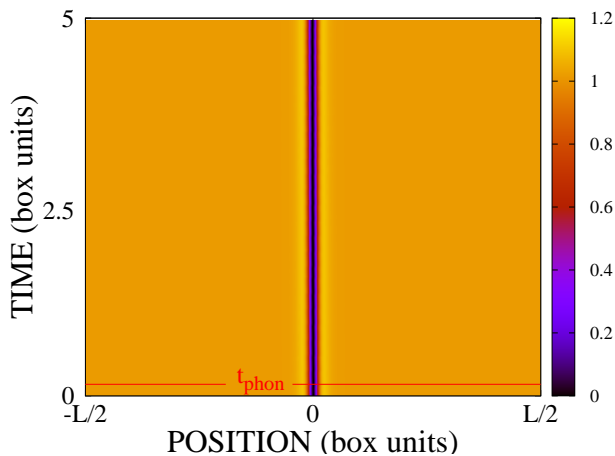}

	\caption{Density as a function of time presented in the  frame rotating with the speed $\pi/L$. 
In the laboratory frame the solitons travel around the circle $17$ times.
	Parameters:  $\lambda=0.01$, $n g_{\rm dd} = 4000$ and $n g_{\rm con} = 0$, as in the Dysprosium like case commented in caption of Fig. \ref{fig:Fig3}.
	The horizontal red line is at time $t_{\rm phon}=0.16$ :  a time at which a single phonon would pass the whole box.
	\label{fig:Fig4}}
	\end{figure}
The presented solutions are only candidates for solitons, having similar jump of the phase as the  Shabat-Zakharov solitons.
The crucial feature of solitons is their ability to propagate without changing their shape. 
To verify if this is the case also here, we were setting as initial states the results of the imaginary time evolution, and 
trace their dynamics obtained via numerical solution of GPE. An example is given in Fig. \ref{fig:Fig4}, where we unrotated the constant speed due to the linear
 dependance of the phase. The evolution time in this figure is about $30$ times longer than a time $t_{\rm phon}$ at which a phonon would 
travel across the whole box.
When testing the system in a wide parameter range and for a  longer evolution time we always observe the dips which propagate without 
noticable distortion.
\section{Two solitons\label{sec:2solitons}}
	\begin{figure}[htb]
	\includegraphics[width=8.7cm,clip]{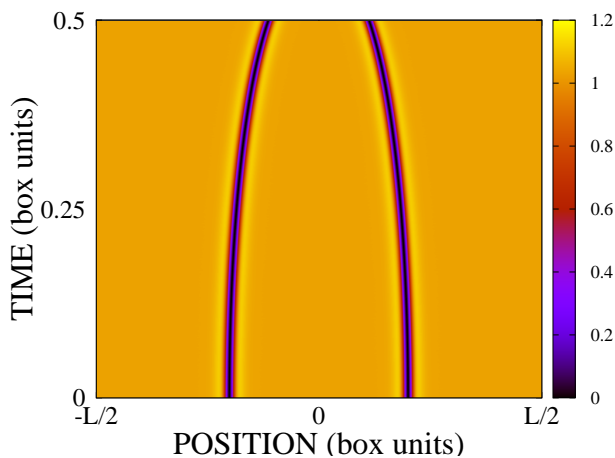}

	\caption{Density as a function of time. The initial distance between solitons is equal to $0.4$.
	Parameters as in Fig \ref{fig:Fig3}.
	\label{fig:Fig5}}
	\end{figure}
The real solitons obey an  equivalent of the superposition principle: although the system is non-linear one can construct many solitonic solution based on a single one. 
As shown in the paper of Shabat and Zakharov \cite{Zakharov1973}  in the contact interacting gas two solitons, when colliding, do not change the shape.
To test if the dipolar solitons can behave similarly, we generate  and propagate pairs of them.
We follow the method described for a single soliton, but this time we enforce two jumps of the phase, at positions $z_1$ and $z_2$, and 
without linear background
	\begin{equation}
		\varphi_2(z) \equiv \frac{\pi}{2} \bb{ {\rm sign} (z-z_1) -   {\rm sign} (z-z_2) }.
		\label{eq:phase2}
	\end{equation}
Having such constraint on the phase we reach with the help of the imaginary time evolution two solitons and then we propagate them using GPE. An example of such 
dynamics is shown in Fig \ref{fig:Fig5}. The immediate conclusion is that the dipolar solitons interact 
, in the example shown they attract each other. Hence, the excitations we study are not solitons in the strict mathematical sense.
Even though, these excitations resemble the shape for a long evolution times, thus we will use the name soliton-like or, for brevity, even the name soliton.

	\begin{figure}[htb]
	\includegraphics[width=8.7cm,clip]{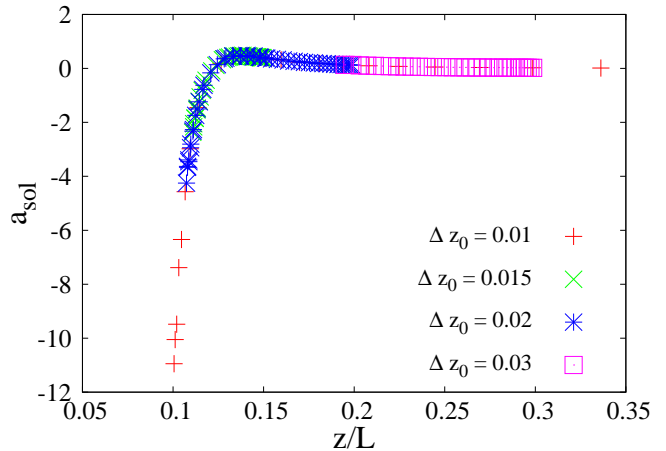}

	\caption{Acceleration versus distance between solitons $a_{\rm sol}(\Delta z_{\rm sol})$ for different initial states.
	Parameters as in Fig \ref{fig:Fig3}.
	\label{fig:Fig6}}
	\end{figure}
To understand the motion of such a pair we monitor the position of one of the solitons versus time: $z_{\rm sol}(t)$. 
Then we numerically compute both the velocity and the acceleration $a_{\rm sol}(t) = \frac{dz_{\rm sol}^2}{dt^2}$. 
Combining these quantities we obtain acceleration versus inter-solitonic distance $a_{\rm sol} ( \Delta z_{sol} )$, where $\Delta z_{sol} = 2 |z_{\rm sol}|$.
We repeat this procedure for many initial states. 
One can see that results of many simulations are lying on a common curve, see Fig. \ref{fig:Fig6}. 
It suggests us to use the following classical mechanics picture, in analogy to many cases for dark solitons in the 
contact interacting gas \cite{Theocharis2005,anglin2000}
we treat each soliton as a point-like (negative) mass obeying the Newton law. 
Assuming $a_{\rm sol} = F_{\rm sol}/m = - \frac{1}{m}\frac{d V_{\rm sol}}{dz} $, 
we can numerically integrate $a_{\rm sol}$ to  extract inter-solitonic potential divided by a mass $V_{\rm sol}/m$.
	\begin{figure}[htb]
	\includegraphics[width=8.7cm,clip]{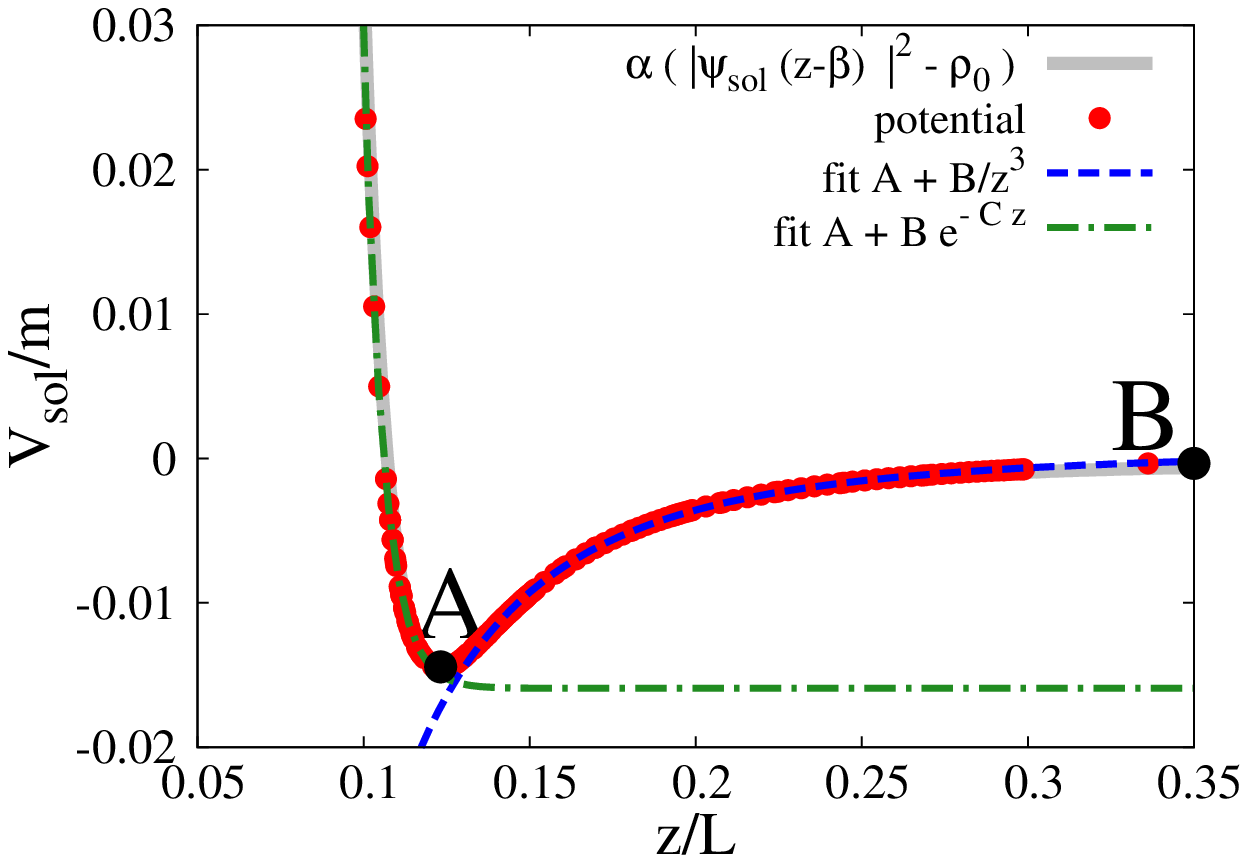}

	\includegraphics[width=3.7cm,clip]{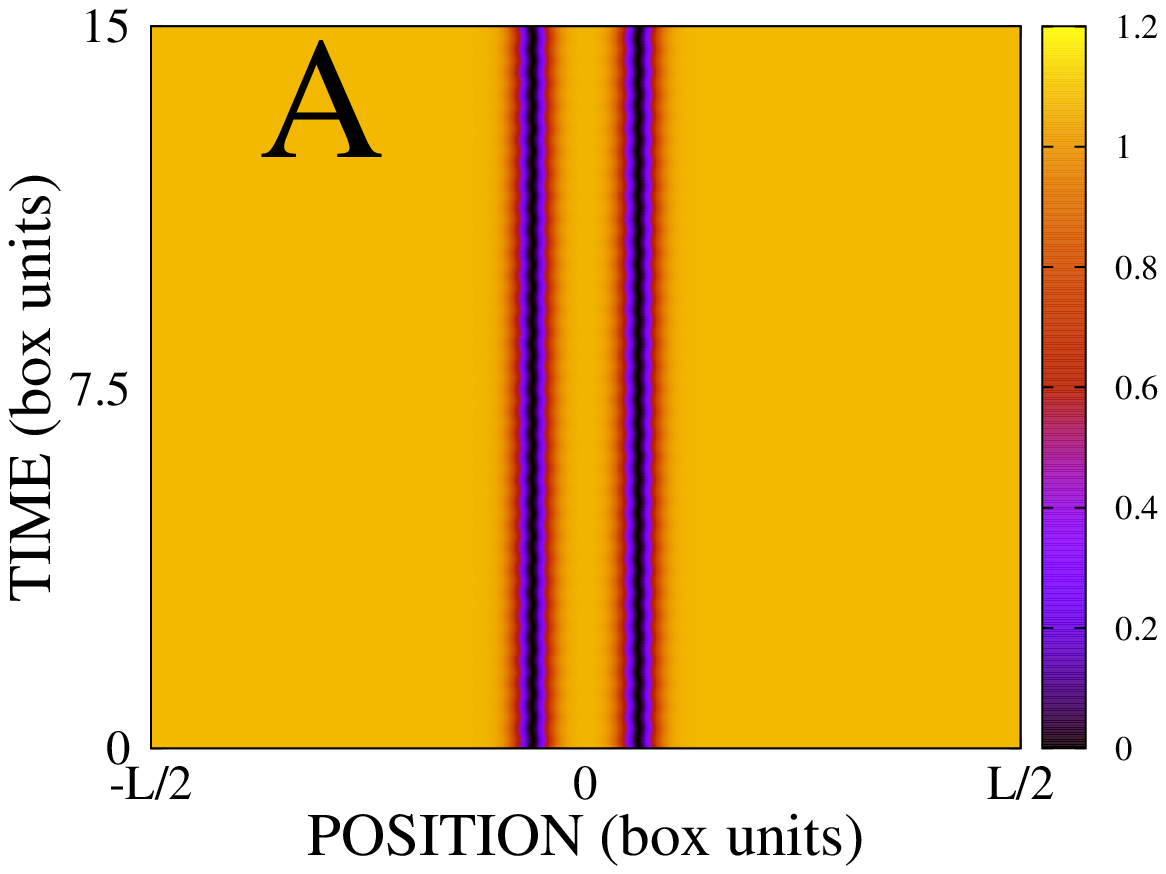}
	\includegraphics[width=3.7cm,clip]{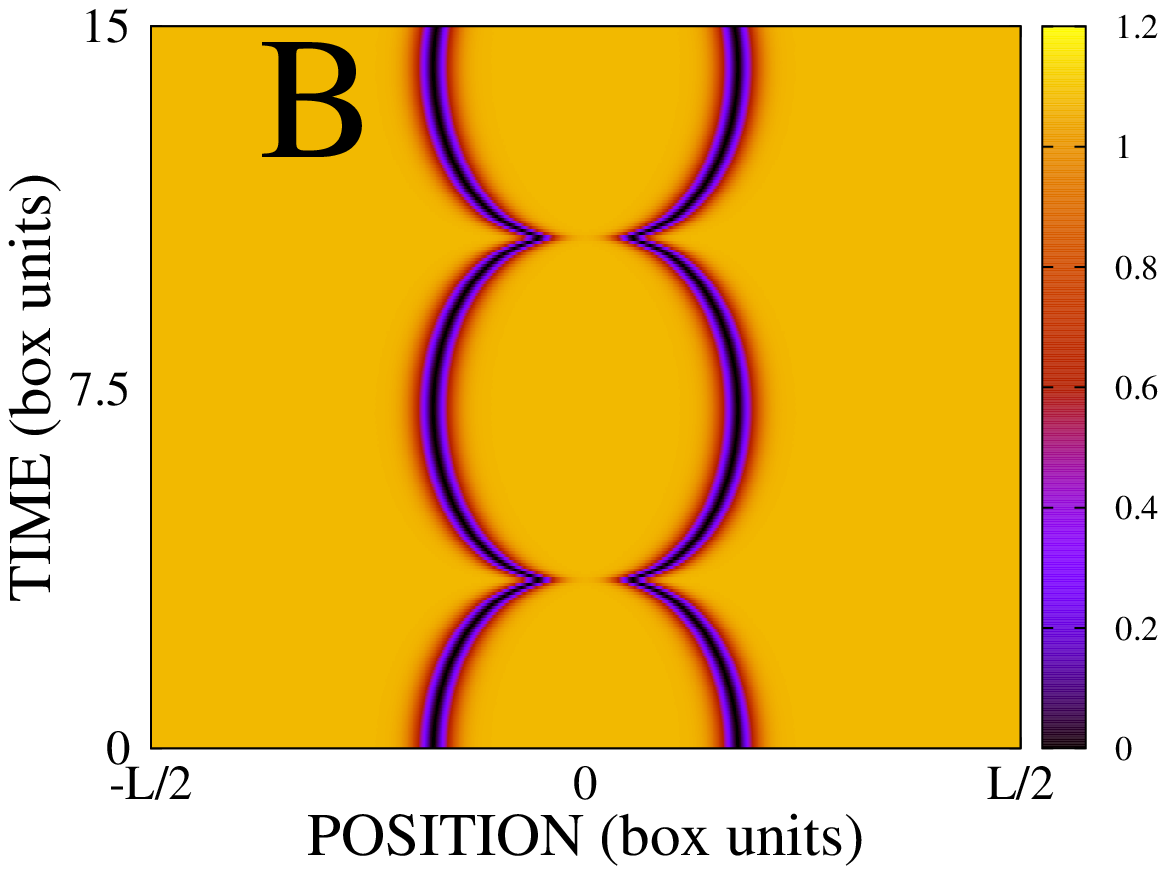}
	\caption{Effective interaction potential between solitons $V_{\rm sol}$ divided by mass in the classical picture.
	The figures in the bottom panel show density versus time for two initial conditions, for the minimum of the potential $\Delta z_{\rm sol} (t=0) = 0.1232 L$ (left) and for $\Delta z_{\rm sol} (t=0) = 0.35 L$ (right).
	Parameters: $\lambda=0.001$, $N g_{\rm dd} = 1000$ and $g_{\rm con} = 0$.
	\label{fig:Fig7}}
	\end{figure}
We show the results for two cases:  in Fig. \ref{fig:Fig7}  for very elongated trap with $\lambda=10^{-3}$ and weak dipolar interaction $Ng_{\rm dd}=1000$
and in Fig.  \ref{fig:Fig7b} for $\lambda=10^{-2}$ and $Ng_{\rm dd} = 4000$.
In both cases the inter-solitonic potential has a repulsive core and the tail scaling with the distance as $1/z^3$.
We tested this classical picture starting again from different initial states and watching their evolution.
Samples of this investigation are given  at the bottom panels in Figs. \ref{fig:Fig7} and  \ref{fig:Fig7b}. 
Two solitons placed close to each other strongly repel each other. 
As the inter-solitonic potential has a local minimum, thus there is a possibility for bound states as shown in the panels marked with the letter A. 
Two far away lying solitons, like in the panel marked with the letter B, they hardly see each other, but after longer time they finally 
start oscillations in the common potential.

In Figs. \ref{fig:Fig7} and \ref{fig:Fig7b} we compare the resulting potential $V_{\rm sol} / m $ just with the rescaled density of a single soliton 
$\alpha \bb{ |\psi (z - \beta)|^2 - \rho_0 }$, where $\rho_0$ is the density far from the soliton, approximately equal to $1/L$
\footnote{We compute $\rho_0$ numerically, it is not fitting parameter.}.
The agreement between solitonic shape and inter-solitonic potential is surprisingly good. 
This indicates the origin of the inter-solitonic potential: each soliton is simply moving in the mean field potential created by its partner. 
As the dark solitons have negative masses, they  are climbing towards the local maxima of the density.
	\begin{figure}[htb]
	\includegraphics[width=8.7cm,clip]{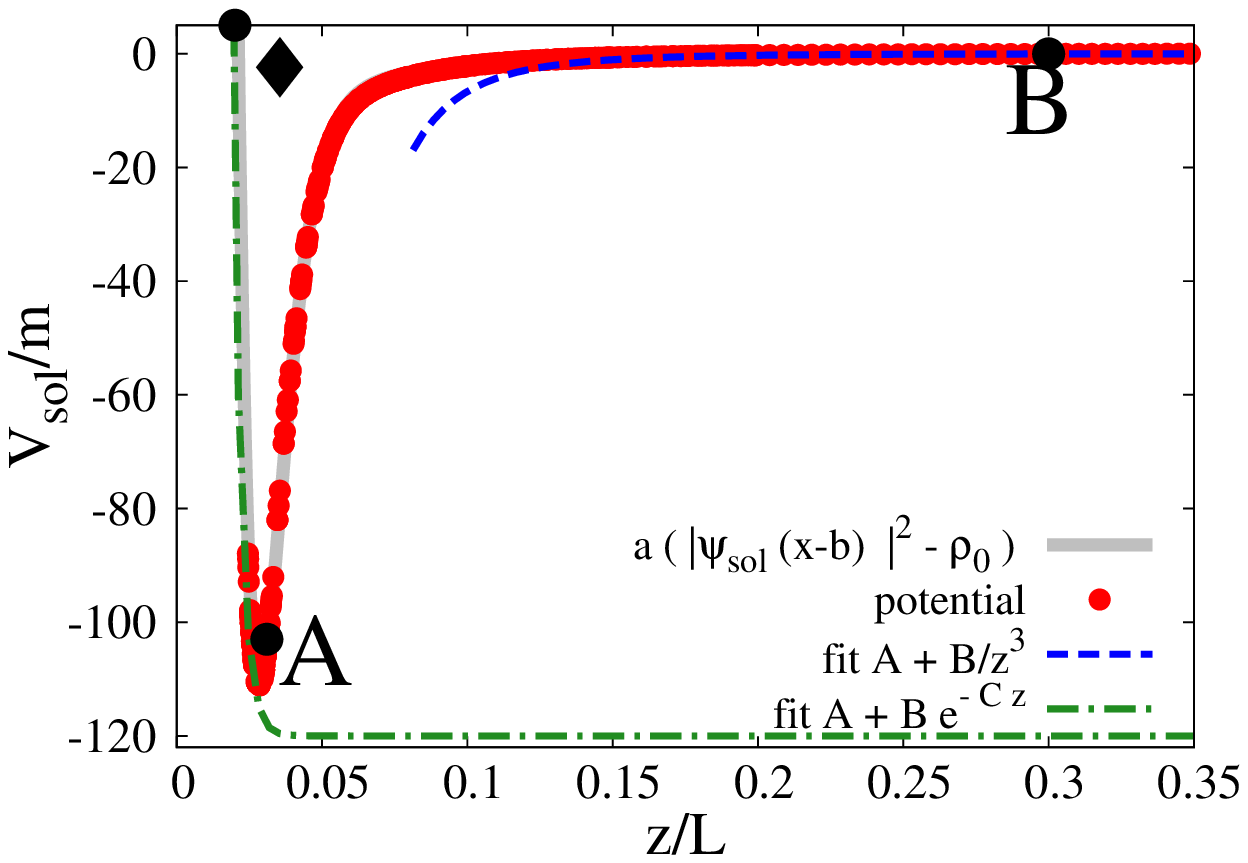}

	\includegraphics[width=2.8cm,clip]{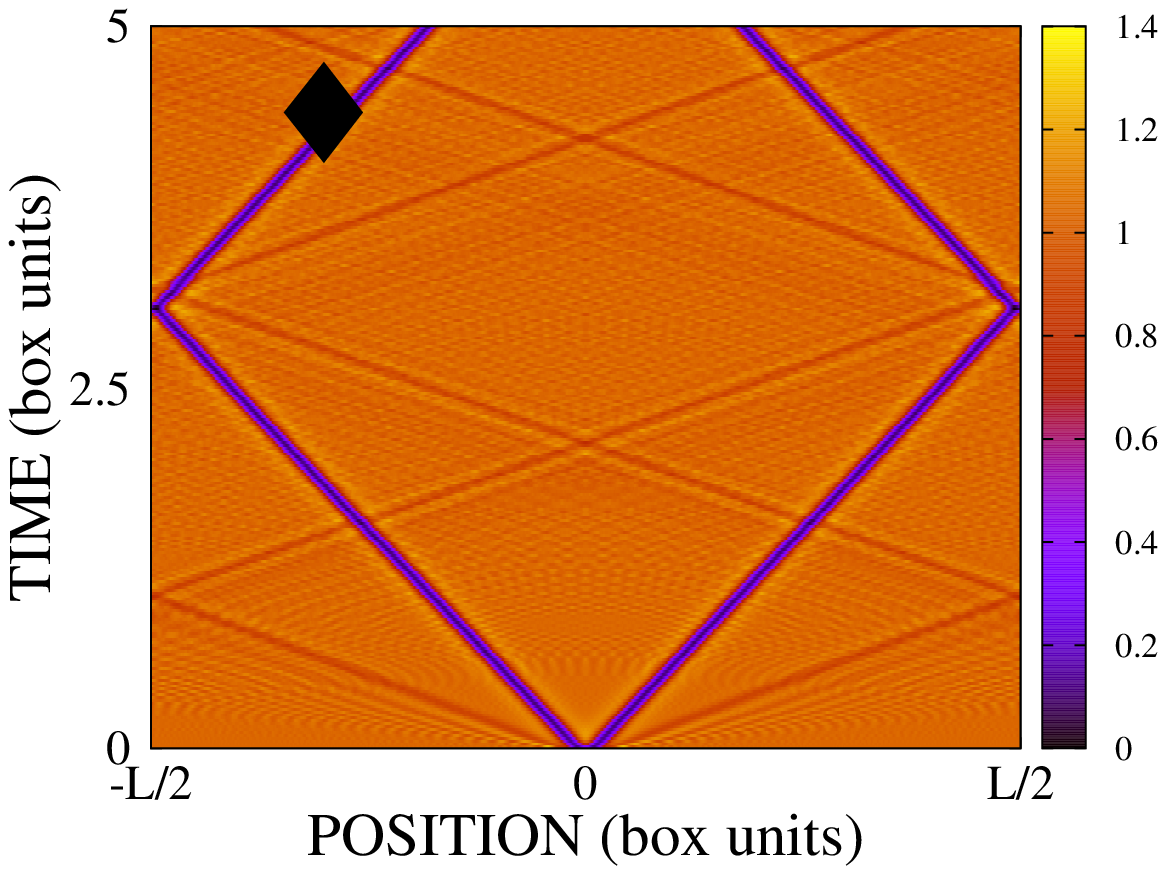}
	\includegraphics[width=2.8cm,clip]{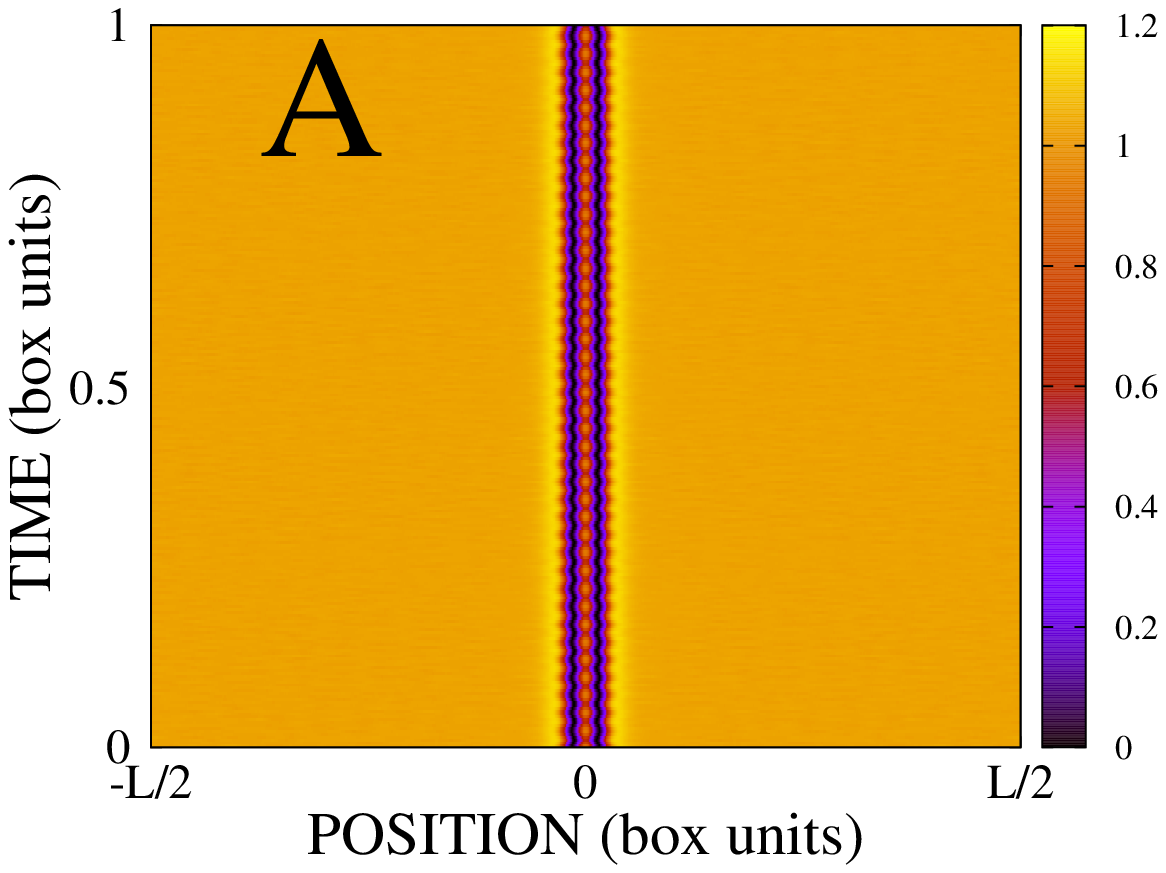}
	\includegraphics[width=2.8cm,clip]{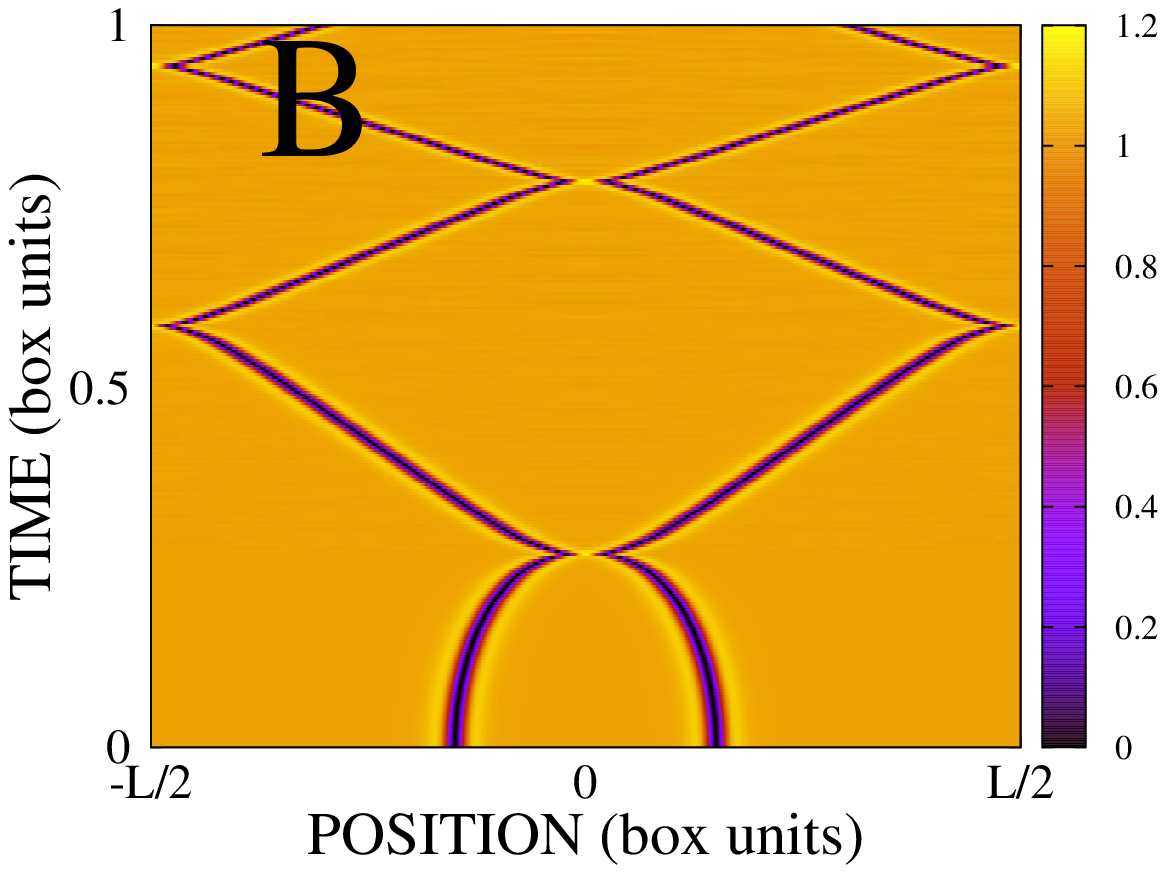}
	\caption{Effective interaction potential between solitons $V_{\rm sol}$ divided by mass in the classical picture.
	The potential consist of repulsive exponentially decaying core and a tail that decays like $1/z^3$.
	The figures in the bottom panel show density versus time for three initial conditions, at the repulsive core
$\Delta z_{\rm sol} (t=0) = 0.022 L$ (left),  close to  the minimum of the potential $\Delta z_{\rm sol} (t=0) = 0.031 L$ (center) and for $\Delta z_{\rm sol} (t=0) = 0.3 L$ (right).
	Parameters as in Fig \ref{fig:Fig3}.
	\label{fig:Fig7b}}
	\end{figure}

\begin{figure}[htb]
	\includegraphics[width=8.7cm,clip]{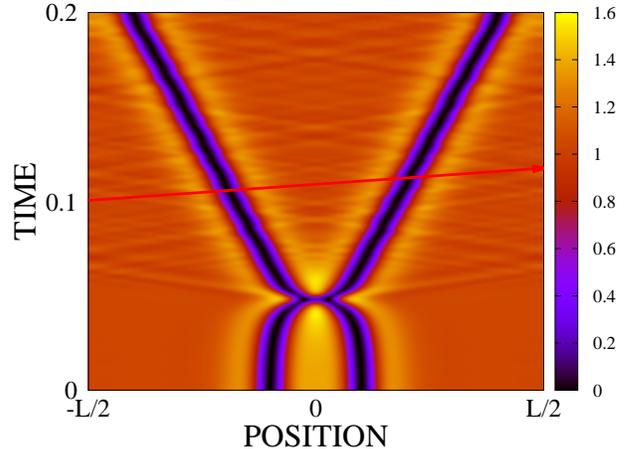}

	\caption{Density as a function of time. The red solid line indicates the speed of sound.
	Parameters: $\lambda=0.1$, $n g_{\rm dd} = 1000$ and $n g_{\rm con} = 0$.
	\label{fig:Fig8}}
	\end{figure}

The attentive Reader could notice some peculiarity in Fig. \ref{fig:Fig8} bottom right panel. 
There, the solitons seem to slightly accelerate. 
This counter-intuitive effect is not a numerical artefact.
To illustrate clearly what is happening we present the evolution for the extreme conditions, $\lambda=0.1$ $ng_{\rm dd}=1000$, in Fig. \ref{fig:Fig8}. Then one can see that the collision is not elastic: colliding solitons radiate phonons. 
In consequence the energy of solitons decays, and the dark solitons are transformed into the gray solitons. 
As in the case with contact interacting gas, the gray solitons  have some characteristic speed.
This extra velocity is thus gained by the solitons due to the inelastic collision.

Hence, the dissipation during the collision leads in fact to acceleration of solitons, as clearly shown in Fig. \ref{fig:Fig8}.
Actually this effect, although not directly described, seems to exist for the  optical solitons with nonlocal nonlinearity studied in \cite{krolikowski2010} (Fig. 3 therein). 
        \begin{figure}[htb]
        \includegraphics[width=8.7cm,clip]{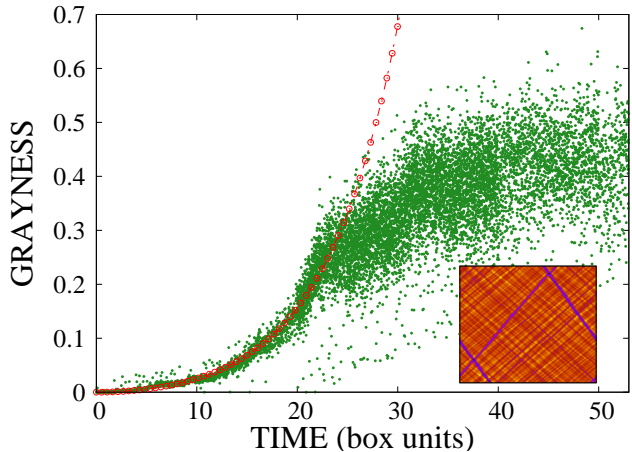} 

        \caption{The grayness $h$ defined in the text  as a function of time. 
	The timescale is so long, that during evolution there occurs hundreds of the collisions between solitons.
        The superimposed red line corresponds to travel at the speed of sound.
	The plot of density vs time corresponds to the evolution at short time window, from $t=40$ to $t=40.01$.
	The red line is a fit in which we invert the energy to grayness $H$ relation $H(E)$ and assume 
	energy $E = E_0 - A {\rm sinh} ( B t)$ with fitting parameters $E_0$, $A$ and $B$, to model the process of the stimulated emission.
	Parameters as in Fig \ref{fig:Fig3}.
        \label{fig:Fig9}}
        \end{figure}
Until collision the movement of a pair of solitons can be captured by a classical model.
Then, collision goes clearly beyond the classical picture: it is an event when the part of the energy is transferred from solitons into initially unoccupied excitations, i.e. phonons. What is the fate of these solitons?
To find the answer we let the system evolve for a long time. During the evolution we monitored minimal density in the system, called here a grayness $H$. 
In the case of a gas with contact interactions only, this parameters can be translated into  the   Shabat-Zakharov soliton speed, equal to 
$v(H) = \sqrt{\frac{H}{\rho_0}} c$, 
and its energy, equal to $E(H) = \frac{\sqrt{2}}{3}\frac{m\hbar}{g_{\rm dd}} c^3  \bb{1  - H L}^{3/2}$ 
\cite{jackson2002}.
The evolution is shown in Fig. \ref{fig:Fig9}.
As we started with two black solitons, initially $H=0$. 
Then, at short times,the grayness  $h$ increases, first quadratically and then exponentially.
At a very long time the quantity $h$ starts to saturate.
We understand the dynamics in terms of interaction between the subsystem (solitons) with an environment (all phonons and other quasiparticles), 
taking the idea from the paper \cite{bach2002}.
Initially the environment is empty: this is the regime of spontaneous emission, where a coupling between subsystem and the unoccupied modes leads to a slow dissipation. Then, we enter the regime of stimulated emission, where the modes after the stage of spontaneous emission are already occupied, and hence the rate of phonon emission rapidly grows. 
If the environment would be infinitely large then eventually we would loose the energy from the subsystem completely. 
However here, we deal with a small environment, in which the phonons always overlap with solitons. This allows for the dynamical equilibration, in which the amount of the energy radiated to phonons is equal to the energy absorbed, as in the qubit interacting with a thermal reservoir.
Indeed we see in the inset of Fig. \ref{fig:Fig9}, that even after a long evolution time the two solitons survive.
The large fluctuations of $H$ are due to overlaps between phonons and solitons, but also due to statistical fluctuation of the flow of the energy between solitons and phonos.
        \begin{figure}[htb]
        \includegraphics[width=4.7cm,clip]{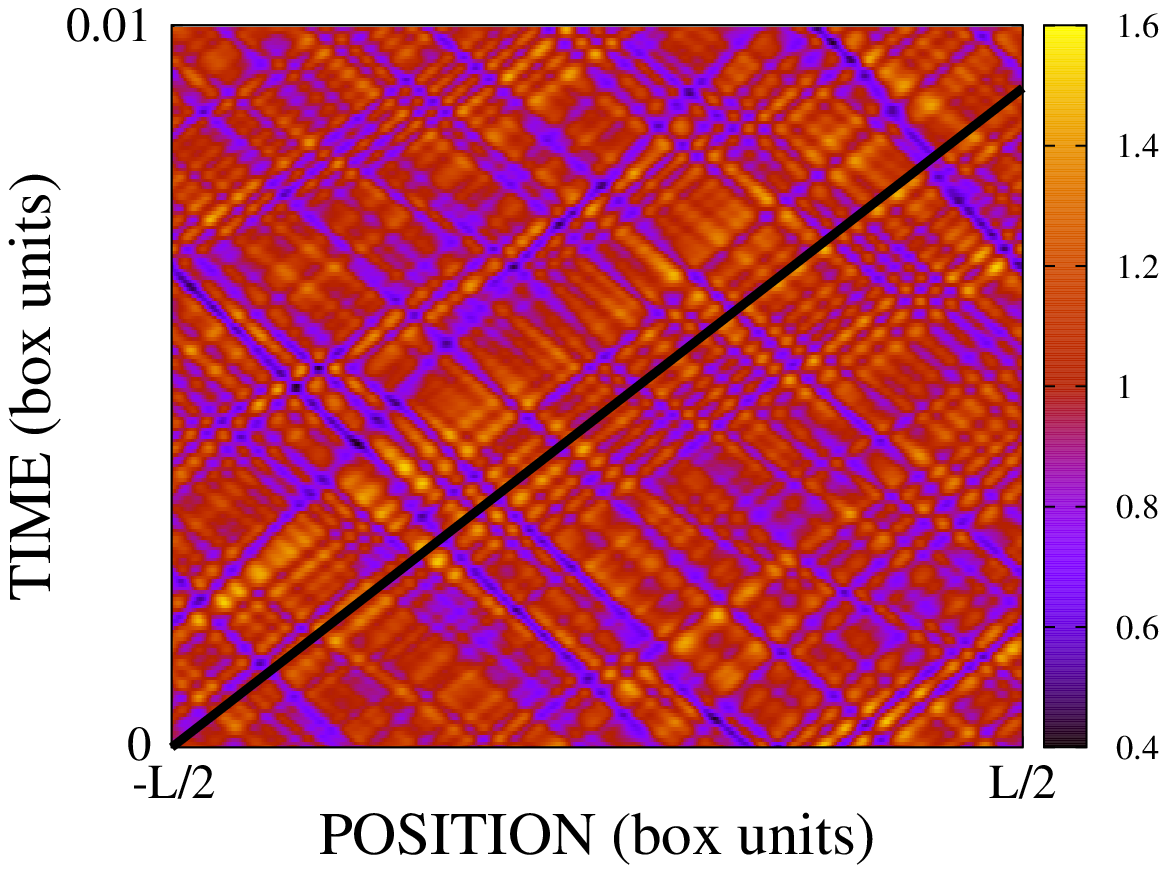} 
        \includegraphics[width=4.7cm,clip]{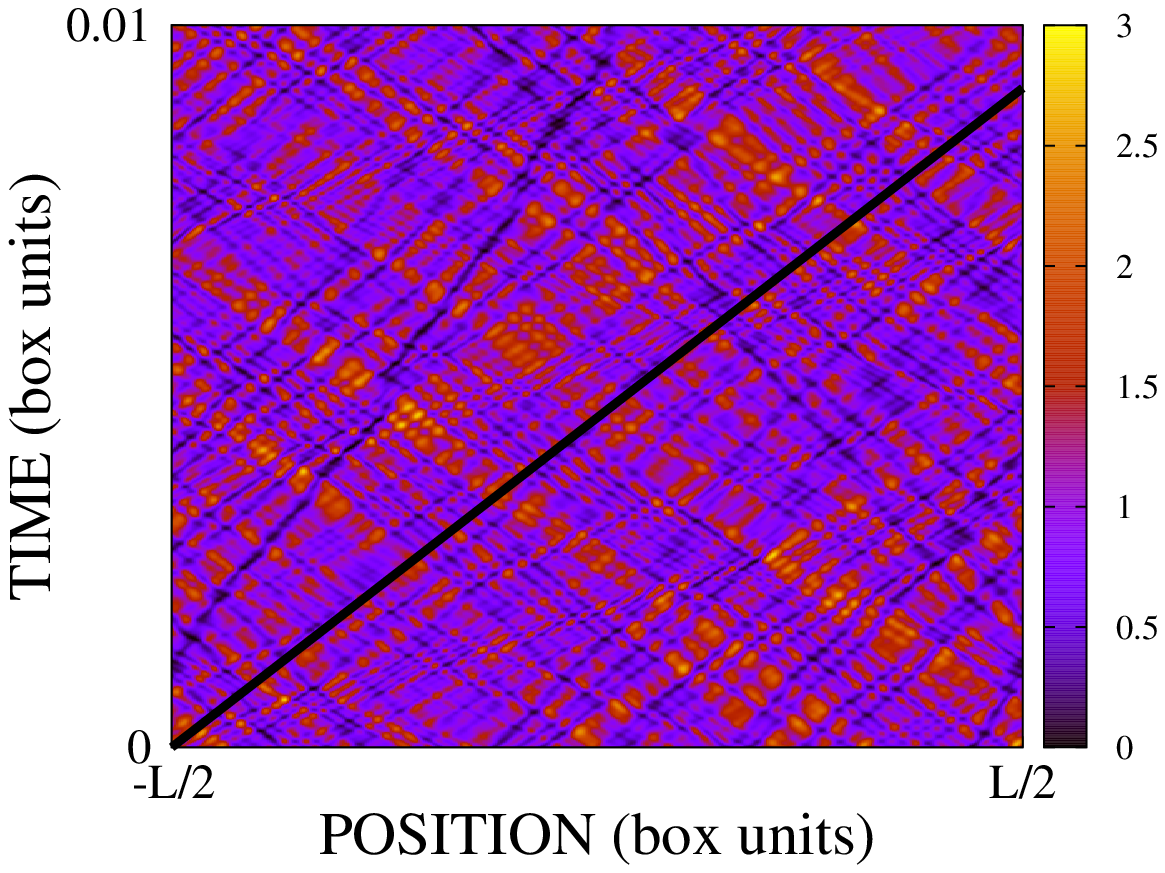} 
        \includegraphics[width=4.7cm,clip]{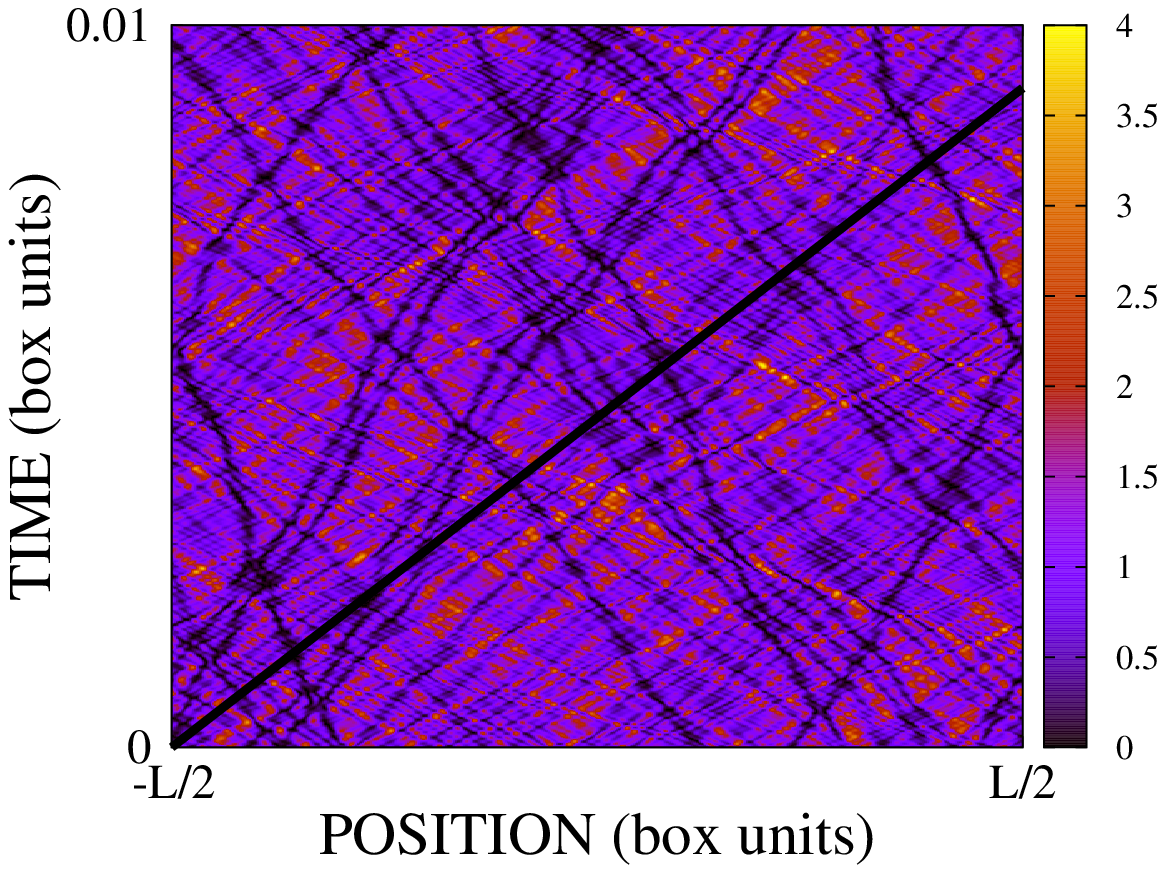} 

        \caption{ Density as a function of time at three different temperatures.
	The thick black line depicts a travel with the speed of sound.
	Parameters: $\lambda=0.01$, $n g_{\rm dd} = 4000$ and $n g_{\rm con} = 0$ and $N=1000$.
        \label{fig:Fig10}}
        \end{figure}
\section{Thermal solitons \label{sec:thermal}}

As mentioned in the introduction, the ultracold gas with contact interaction possesses not only phonons but also dark solitons \cite{karpiuk2012}. 
These solitons are the elementary excitations forming the second branch of excitations derived in 1963 by Elliot Lieb \cite{lieb1963II}. 
We would like to check if non-Bogoliubov excitations exist also in a quasi 1D dipolar gas.
To fully verify it, one should solve a many-body problem, an equivalent of the Lieb-Liniger model for a dipolar gas. 
This task is definitely beyond this paper. 
Instead, still relying on the mean field model, we can at least check if the solitons, as candidates for "type II" excitations in a dipolar gas, appear spontaneously at finite temperatures and if their number obeys the thermal statistics.
At the first glance, the presence of such solitons seem unlikely: the system is not integrable and we have already shown that the dipolar soliton-like excitations are not robust, as they collide inelastically. 
On the other hand, we also  demonstrated  that  two solitons immersed in the gas full of phonons can stay in the dynamical equilibrium. The latter observation encourages us to look also at the thermal equilibrium.

As in \cite{karpiuk2012}, we use the classical field approximation (CFA), in which one replaces the quantum field operator $\hat{\Psi}(z)$ with the classical field $\Psi (z)$.
This method is commonly used at zero-temperature, where  $\Psi (z)$ is interpreted as the macroscopically occupied (by all particles) single particle orbital, called the GPE wave function.
The idea of using the same replacement to describe a gas at non-zero temperature relies on the fact that for the degenerate gas there are many substantially occupied single-particle orbitals, which sum up to a classical field $\Psi(z)$.

To compute the values of target variables averaged over a statistical ensemble one performs averaging over
the appropriately chosen set of the classical fields $\Psi(z)$ or/and time-averaging over the fields propagated with the help of the GPE equation or its generalization.

There is variety of different classical field methods devised to describe ultracold gases \cite{alice2013}.
Here we use a simple version described in \cite{Brewczyk2007}. 
First, we generate many initial states using a method based on the Bogoliubov description.
As our method of generating thermal samples is not sufficiently good, we evolve the field with the help of GPE, we trace the system for sufficiently long time and compute time averages. 

More precisely, in each realization, the initial wave-function reads
	\begin{equation}
		\Psi(z) = \frac{1}{\sqrt{L}} \sum_{ - k_{\rm max} < k< k_{\rm max} } \alpha_{k}  e^{- i k z},
	\end{equation}
where the wave vector is $k = 2\pi l/L$ with $l$ an integer. The sum over wave-vectors is limited by the 
 $k_{\rm max} = \sqrt{c k_{\rm B} T  + \mu   }$ with $c = 0.29$, where $T$ is the temperature and $\mu = n V_{\rm eff}(k=0)$ is the chemical potential. 
The cut-off $k_{\rm max}$  is chosen such that with the classical field method one would recover the quantum results at least for the ideal gas
\cite{emilia_pra_2009,Witkowska2010671}.
The expansion coefficients $\alpha_k$ should be drawn from the thermal classical distribution. Even on the classical level this drawing can be a cumbersome task. 
The details of our procedure of drawing initial set of $\alpha_k$ are given in the Appendix.
As the procedure gives unsatisfactory effect (especially for higher temperatures), we have to propagate each guessed classical field over a long time to allow  for thermalization.
Finally, after this auxiliary stage of thermalization, in the next stage we further propagate the state to perform averaging over time.
The short-time windows of the evolution during the last stages are illustrated in Fig \ref{fig:Fig10}.
We see there three examples of the evolution of the density of the gas, with the fraction of the $0$-momentum mode  from $0.68$ (top panel), through  $0.17$ (middle panel), $0.08$ (bottom panel).

The speed of sound is depicted with the solid black line. 
In the middle and the bottom panels one can see dips in the density, propagating much slower than the speed of sound. 
We checked that these dips were stable even at long timescales. Thus we expect that they are dark solitons.
        \begin{figure}[htb]
        \includegraphics[width=8.7cm,clip]{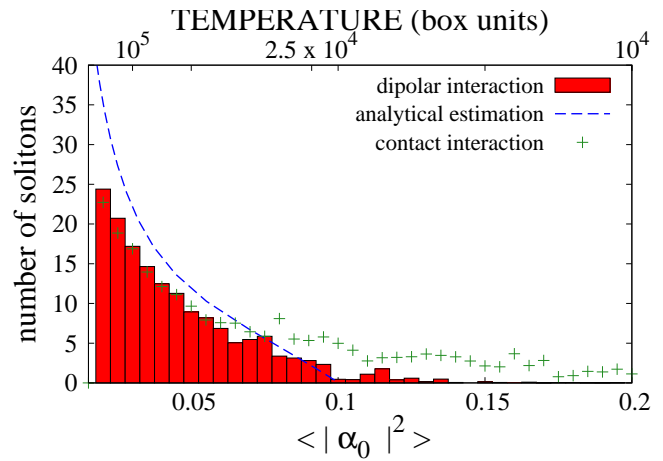} 

        \caption{Number of dark solitons versus fraction of atoms in the $0$-momentum mode.
	Histogram corresponds to the numerically treated dipolar case  with parameters
	$\lambda=0.01$, $n g_{\rm dd} = 4000$ and $n g_{\rm con} = 0$.
	The green points are computed analogously to the dipolar case, but for $n g_{\rm con}=12000$ and  $g_{\rm dd} = 0$.
	Finally the blue dashed line is the expected number of solitons assuming the equipartition of energy and assuming that
dark solitons as quasi-particles, namely number of solitons is approximated with $k_{\rm B}T/E_{\rm sol}$, where $T$ is the estimated temperature and $E_{\rm sol}$ is the estimated energy of the dark soliton.  The details of the estimation are in the text.
	In each case we assumed $N=1000$ atoms. The results for the dipolar gas and the gas with contact interaction comes from time averaging performed on $1000$ realizations of the classical field, each drawn as described in the Appendix.
        \label{fig:Fig11}}
        \end{figure}

To convince ourselves we prepare the following test: for each thermal sample we performed a long evolution, during which we average over time the
fraction of the least energetic $0$-momentum mode $\meanv{|\alpha_0|^2/N}$, the fluctuation of this fraction $\Delta \bb{|\alpha_0|^2/N}$  and  the number of the deepest dips, with the density below $0.05/L$.
The last quantity, called the number of dark solitons, is presented in Fig. \ref{fig:Fig11}.
We estimated the temperature of each sample using formulas for average number of motionless particles, 
i.e. $\meanv{|\alpha_0|^2/N}$ , for the ideal gas \cite{wilkens1997}.
As the crosscheck, we compute what would be fluctuation of  the latter average in the ideal gas and compare it with the previously numerically found $\Delta\bb{|\alpha_0|^2/N}$. The agreement was within a few percent, hence we concluded that: 1) the interactions in the gas are still so small that the statistics seems close to the ideal gas case and 2) our estimation for temperature is reasonable within a few percent.
We also estimated the energy of dark solitons, relying on the formulas for the gas with contact interactions only \cite{jackson2002}
as done in the previous sections. This gives the estimation for the energy of the dipolar dark soliton 
$E_{\rm sol} = \frac{4\sqrt{2}}{3} \sqrt{3 n g_{\rm dd}}$ (in box units).
Finally, having the estimated energy of the dark soliton, and estimated temperature of each sample we computed the estimated number of dark solitons $k_{\rm B} T/E_{\rm sol}$, assuming their equipartition as it should be for quasiparticles. 
This expected number of solitons is given by the blue dashed line in Fig. \ref{fig:Fig11}.
We also repeat all the steps for the gas with the contact interaction only.

\section{Conclusions}
We investigated the dipolar gas confined in the potential having the shape of a long tube, with the dipole moments polarized perpendicular to the long axis. In this geometry we find robust localized excitations, very similar to the Shabat - Zakharov solitons appearing in the gas with contact interaction only.
In contrast to the contact interacting case, pairs of these soliton-like dipolar excitations mutually interact, and are able to form two-soliton bound states.
Moreover, the collisions between such pairs are not elastic: during collision they emit phonons, hence they loose energy and because of this they accelerate. After thousands of subsequent collisions they can reach dynamical equilibrium with the already emited quasiparticles.
Finally we show that these excitation appear spontaneously at finite temperatures and their number obeys thermal statistics.
The latter observation is closely related to the contact interacting case, where the appearance of the dark solitons at the non-zero temperatures is an evidence that they are the non-Bogoliubov elementary excitations, predicted by Elliot Lieb in 1963.

There are many open questions concerning these excitations. 
One should check their stability in the real three dimensional calculation. It is interesting if they exist  also in the roton regime. 
There is the question of the effect of a harmonic trap along the long axis (here the box with periodic boundary was assumed).
One needs also benchmarks from other methods used for the finite-temperature simulations, as they are quite developed for the dipolar gases \cite{ticknor2012,bisset2012,blakie2009,bisset2011}.
 Probably the most important task would be a verification if these  excitations are really elementary and if they are present in the many-body model.

\begin{acknowledgments}
We are grateful to T. Karpiuk, T. Sowi\'nski, P. Deuar and M. Gajda for valuable discussions. 
The work was supported by the (Polish) National Science Center Grant No. DEC-2012/04/A/ST2/00090.
\end{acknowledgments}

\section{Appendix: Generating thermal states within classical field approximation}
In the Bogoliubov method one approximates the grand canonical Hamiltonian with $H - \mu N = \sum_k E_k \hat{b}_k^{\dagger} \hat{b}_k$, 
where $E_k =\sqrt{\frac{k^2}{2 m} \bb{ \frac{\hbar^2k^2}{2m} + 2 n \tilde{V}_{\rm eff} (k)  }} $ is the Bogoliubov spectrum and the operators 
$\hat{b}_k$ and $\hat{b}_k^{\dagger}$ are the annihilation and creation bosonic ladder operators for quasiparticles, respectively.
The operator $\hat{b}_k$, annihilating a quasiparticle with quasi-momenta $k$ is related to the bosonic operators $\hat{a}_k$ annihilating 
a particle with a given momenta $k$ via relation:
	\begin{equation}
		\hat{b}_k = \hat{a}_k u_k - \hat{a}_k^{\dagger} v_k,
		\label{eq:cfa}
	\end{equation}
where $u_k = \sqrt{ \frac{\frac{k^2}{2 m}  +  n V_{\rm eff}(k=0) }{2E_k} + \frac{1}{2}}$ and $v_k = -\sqrt{ \frac{\frac{k^2}{2 m}  +  n V_{\rm eff}(k=0) }{2E_k} - \frac{1}{2}}$.
In the classical version, we replace $\hat{a}_k$, $\hat{a}_k^{\dagger}$, $\hat{b}_k$ and $\hat{b}_k^{\dagger}$  with the complex numbers denoted as $\alpha_k$, $\alpha_k^{*}$, $\beta_k$ and $\beta_k^{*}$, respectively.
We introduce the classical probability distribution 
$P (\beta_k, \beta_{k}^*) = \frac{1}{Z} \exp\bb{ - (\sum_{ - k_{\rm max} < k< k_{\rm max} } E_k |\beta_k|^2  )}$ with the partition function defined such that $1 = \int d^2 \alpha_{-k_{\rm max} }\int d^2 \ldots \int d^2 \alpha_{k_{\rm max} } P$.
Within this approximation the number of quasiparticles with momenta $k$ is given by the exponential distribution with the parameter $k_{\rm B} T/E_{k}$.
To generate the initial state foe a given temperature and for a given number of atoms $N$ we follow the steps:
\begin{itemize}
	\item[1)] For $k\neq 0$ we draw the number of quasiparticles $n_k = |\beta_k|^2$  using the exponential distribution.
	\item[2)] For $k\neq 0$ we draw a phase  $\phi_k$ with the uniform distribution on the interval $[-\pi, \pi)$.
	\item[3)] We construct the amplitude $\beta_k= \sqrt{n_k} e^{i \phi_k}$.
	\item[4)] We compute the amplitudes $\alpha_k = u_k \beta_k + v_k \beta_{-k}^{*}$.
	\item[5)] We compute the amplitude of the $0$-momentum mode, $\alpha_0 = \sqrt{N - \sum_{k\neq 0} |\alpha_k|^2}$.
	\item[6)] We construct the classical field according to Eq. \eqref{eq:cfa}.
\end{itemize}

\end{document}